\documentclass[%
 reprint,
 amsmath,amssymb,
 aps,
]{revtex4-1}

\usepackage{graphicx}
\usepackage{dcolumn}
\usepackage{bm}
\usepackage{comment}

\usepackage[mathlines]{lineno}

\usepackage{multirow}
\usepackage{array}
\usepackage{gensymb}

\begin{document}

\preprint{APS/123-QED}

\title{Controlling branching angle of waveguide splitters based on GRIN lenses }

\author{S. Hadi~Badri$^1$}
\author{M. M. Gilarlue$^1$}
\author{S. G. Gavgani$^2$}

\affiliation{$^1$Department of Electrical Engineering, Sarab Branch, Islamic Azad University, Sarab, Iran}
\affiliation{$^2$Department of Electrical Engineering, Azarbaijan Shahid Madani University, Tabriz, Iran}

\date{\today}

\begin{abstract}
Designing beam splitting structures with wide branching angles is of great significance. The branching angle of conventional Y-junctions is limited. In this paper, we investigate the possibility of utilizing gradient index (GRIN) lenses with two focal points such as the generalized Maxwell’s fisheye (GMFE) and Eaton lenses in controlling the branching angle of power splitters. The GMFE lens can provide a wide range of branching angles, however, we present only splitting angles of 25$^{\circ}$, 45$^{\circ}$, and 65$^{\circ}$. Furthermore, we propose a 90$^{\circ}$ splitter structure by employing the Eaton lens. We evaluate the performance of the proposed power splitters by ray-tracing and full-wave finite element method. While GRIN lenses provide a broad range of splitting angles, they require isotropic metamaterials to implement high refractive indices at the center of these lenses.
\end{abstract}


\maketitle

\section{\label{intr}Introduction}
Splitting and combining the optical signals in photonic integrated circuits (PICs) rely on power splitters. Power splitters or Y-branch structures are the key elements in Mach-Zehnder interferometers, optical switches, optical phase arrays, mode multiplexers, semiconductor lasers, samplers, logic gates, and hybrid-integrated optical transceivers \cite{wang2002optimal, hatami1994new, li2013compact,safavi1993wide}. The branching angle of conventional power splitters is usually lower than 12$^{\circ}$ \cite{huang2010low}. The conventional power splitters suffer from severe radiation loss as the branching angle increases. Reducing radiation loss can be achieved by decreasing the branching angle and increasing the length of the splitting structure, resulting in a larger footprint \cite{wang2011design}. Various methods have been studied to control the branching angle of splitters. T-junctions have been implemented based on the left-handed properties of the metamaterial \cite{chen2003t,caloz2001full}. Transformation optics (TO) offers unprecedented control over the flow of electromagnetic fields \cite{leonhardt2009transformation}. Beam splitters with various branching angles have been designed by TO \cite{rahm2008transformation,rahm2008optical,wu2016three,viaene2017mitigating,viaene2016transforming}. These designs are usually implemented by anisotropic metamaterials.
Recent advances in metamaterials and nanofabrication techniques have turned the attention of researchers to the classical GRIN lenses such as Maxwell’s fisheye \cite{gilarlue2019photonic,badri2019multimode}, Luneburg \cite{sayanskiy2017broadband,quevedo2018glide}, and Eaton \cite{badri2019low,badri2019polymer} lenses. In this paper, we present novel optical power splitters by employing GRIN lenses with dual focal points. These lenses provide a flexible structure which could be considered as a proper choice for a wide range of splitting angle. Splitters with branching angles of 25$^{\circ}$, 45$^{\circ}$, and 65$^{\circ}$ are presented by the GMFE lens while a 90$^{\circ}$ branching angle is designed by the Eaton lens. The cost of this fle
xibility is a high refractive index at the center of the lens which requires isotropic metamaterials in the fabrication process. On the other hand, the splitters designed by TO require anisotropic metamaterials. The proposed splitters are evaluated by ray-tracing and full-wave simulations. Recently, the performance of the GMFE as a beam splitter has been studied in the GHz range \cite{lei2017generalized} where a point source is used to evaluate the performance of the lens. A point source can only be a valid estimate of the performance of the GMFE lens as waveguide splitter if only the radius of the lens is considerably larger than the width of the waveguides. Therefore, we take a different approach by using an array of point sources in ray-tracing simulations to achieve a more reliable result. Moreover, conditions for maximum transmission and minimum reflection are discussed.

\section{\label{Spli}GMFE lens as splitter }
\begin{figure}[!t]
\centering
\includegraphics[width=1.0\linewidth]{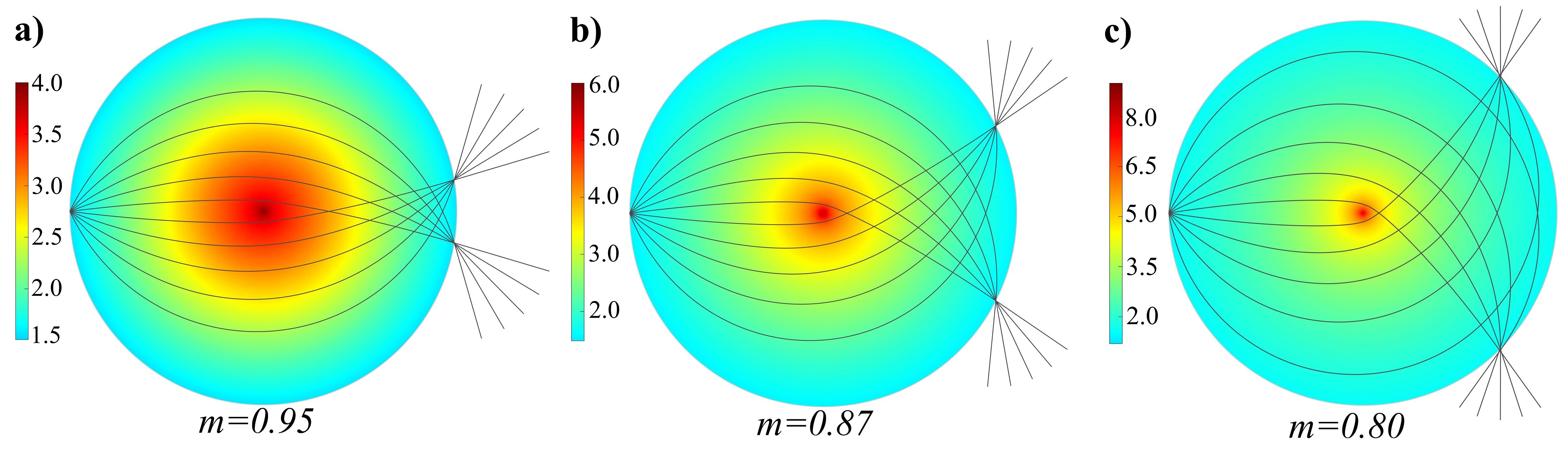}
\caption{Ray trajectory based on a point source for GMFE lenses with a) $m=0.95$, b) $m=0.87$, and c) $m=0.80$.}
\label{fig:Rays}
\end{figure}

Maxwell’s fisheye (MFE) lens is a circular lens of radius $R_{lens}$ with refractive index profile of

\begin{equation}
\label{eq:fisheye_n}
\begin{split}
   &n_{lens}(r)  = \frac{2 \times n_{edge}}{1 + (r/R_{lens})^2}{\rm{    }},{\rm{    }}(0 \le r \le R_{lens})
\end{split}
\end{equation}

where $r$ is the radial distance from the center of the lens, and $n_{edge}$ is the refractive index of the lens at its edge. When rays leave a point source on the edge of the MFE lens, they are focused to the diagonally opposite point of the lens. The MFE lens can be generalized to have two focal points. The refractive index profile for this lens is \cite{tyc2011absolute}

\begin{equation}
\label{eq:gen_fisheye_n}
\begin{split}
   & n_{lens}(r) = \frac{2 \times n_{edge}}{{(r/R_{lens})}^{1 - m} + {(r/R_{lens})}^{1 + m}}{\rm{    }},{\rm{    }}(0 \le r \le R_{lens})
\end{split}
\end{equation}

where $m$ is a variable parameter. Eq.~(\ref{eq:gen_fisheye_n}) reduces to Eq.~(\ref{eq:fisheye_n}) for $m=1$. For $0.5\le m \le 1$, the GMFE lens splits the rays emitting from a point source on its edge into two points on its edge. For $m=0.5$, the lens focuses the rays back to its source after a single revolution around the center \cite{eskandari2019elliptical}. For three values of m, the ray diagrams are shown in Fig~\ref{fig:Rays}. By decreasing m, the angular separation between the focal points as well as the refractive index of the lens increases. The refractive index of the GMFE approaches to infinity at the center. Therefore, we limit the refractive index profiles in the figures presented in this paper in order to make them more distinguishable for the readers. In subsection ~\ref{GO}, we present ray-tracing calculation results for three branching angles. And in subsection `\ref{WO}, the two-dimensional (2D) full-wave simulations are presented for the same structures.

\subsection{\label{GO} Geometrical Optics }
\begin{figure}[!t]
\centering
\includegraphics[width=1.0\linewidth]{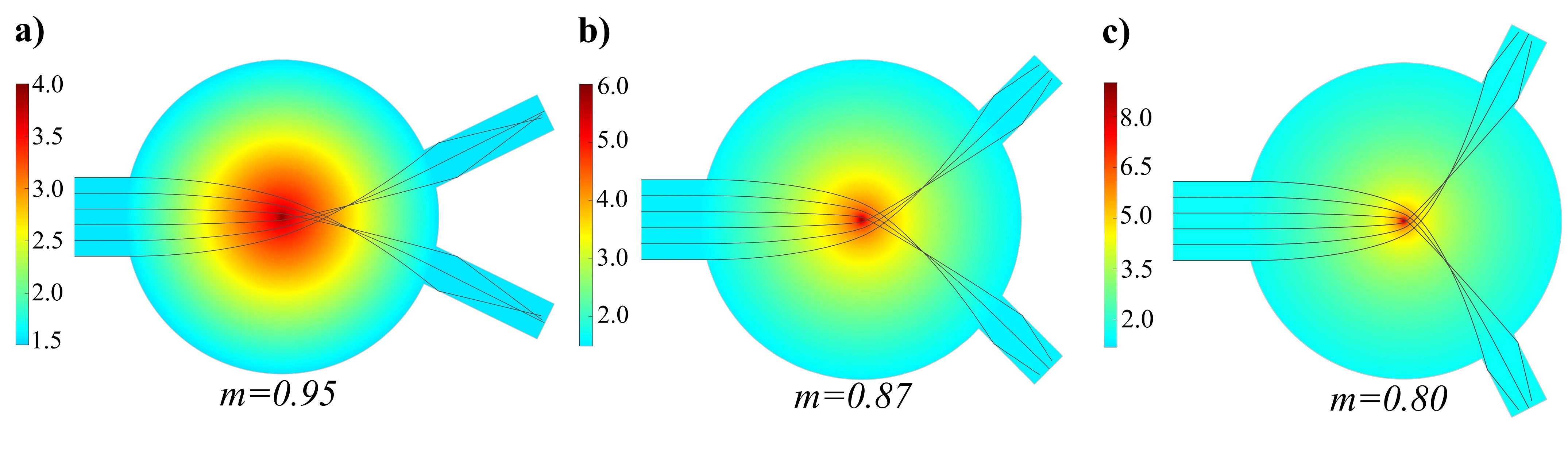}
\caption{Three splitters based on GMFE lenses with branching angles of a)$25^{\circ}$ b)$45^{\circ}$, and c)$65^{\circ}$ determined by an array of point sources.}
\label{fig:Rays_GMFE}
\end{figure}

In this section, ray-tracing calculations are performed with Comsol Multiphysics to validate the proposed splitters. Three splitters with branching angles of $25^{\circ}$, $45^{\circ}$, and $65^{\circ}$ are presented by the GMFE lens in Fig~\ref{fig:Rays_GMFE}. In this study, we suppose a $250 nm$-thick $SiN$ guiding layer which is surrounded by $SiO_2$ substrate and upper air cladding. The effective refractive index of this waveguide is about 1.57. In two-dimensional (2D) simulations, we consider a waveguide with $n_{core}=1.57$ which is surrounded by air. For reducing reflection from the interface of the waveguide and lens, the refractive indices of the waveguide’s core ($n_{core}$) and the edge of the lens should be equal. Therefore, minimizing the reflection is achieved by considering $n_{edge}=1.57$ in Eq.~(\ref{eq:gen_fisheye_n}). The radius of the lens is $R_{lens}=4 \mu m$ while the width of the source waveguide is $2\mu m$. And the width of branching waveguides is $1\mu m$ . A point source may be a reasonable approximation for an input waveguide with a very narrow width. Therefore, in Fig~\ref{fig:Rays_GMFE}, an array of point sources located in the core of the waveguide is used to accurately evaluate the performance of the GMFE lens as a splitter. It should be noted that, while the refractive index profiles of the lenses are the same in Fig~\ref{fig:Rays} and ~\ref{fig:Rays_GMFE}, the splitting angles differ due to the difference between the light sources. For instance, the branching angle in Fig~\ref{fig:Rays}(c) is 90$^{\circ}$ while it is 65$^{\circ}$ in Fig~\ref{fig:Rays_GMFE}(c). Therefore, a single point source cannot be used to determine the branching angle of the lenses, and an array of point sources in the core of the waveguide should be employed to design waveguide splitters.

As seen in Fig~\ref{fig:Rays_GMFE}, the rays are limited to a small area of the lens, therefore, the lens can be truncated without any degradation of its performance. This truncation reduces the footprint of the splitters considerably. Fig~\ref{fig:Rays_GMFE}(a) corresponds to a truncated lens with $m=0.95$ where the splitting angle is $25^{\circ}$. The splitting angle can be increased by reducing m. The splitting angle is $45^{\circ}$ and $65^{\circ}$ for $m=0.87$ and $0.80$, respectively.

\begin{figure}[!t]
\centering
\includegraphics[width=1.0\linewidth]{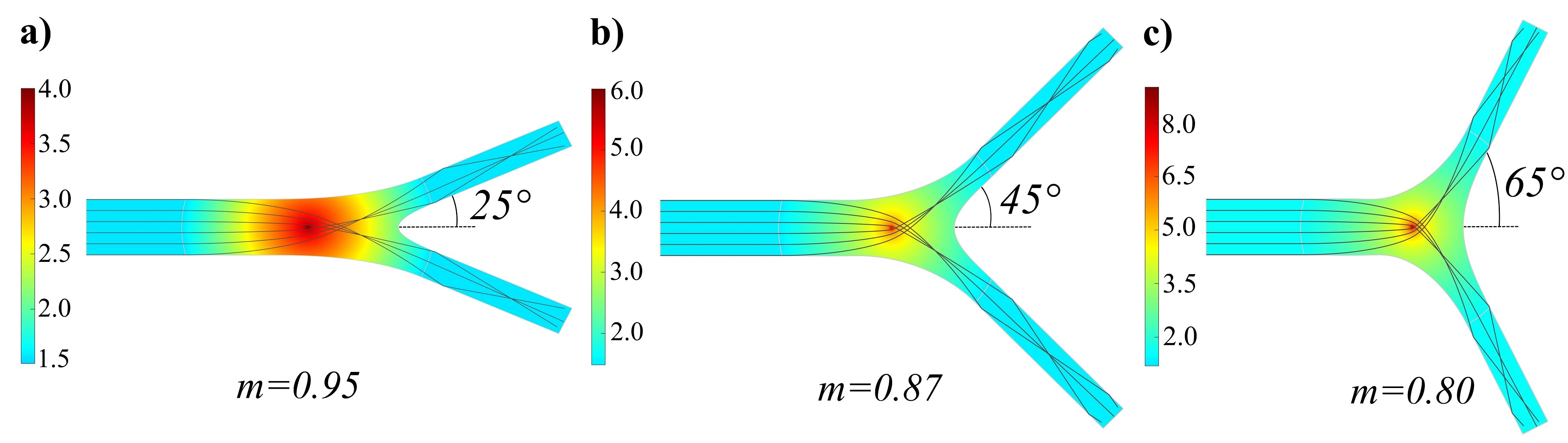}
\caption{The circular lenses of Fig~\ref{fig:Rays_GMFE} are truncated to reduce the footprint of the beam splitters. The branching angles are a) $625^{\circ}$ b) $45^{\circ}$, and c) $65^{\circ}$..}
\label{fig:Rays_truncated}
\end{figure}

\subsection{\label{WO} Wave Optics }
\begin{figure}[!t]
\centering
\includegraphics[width=1.0\linewidth]{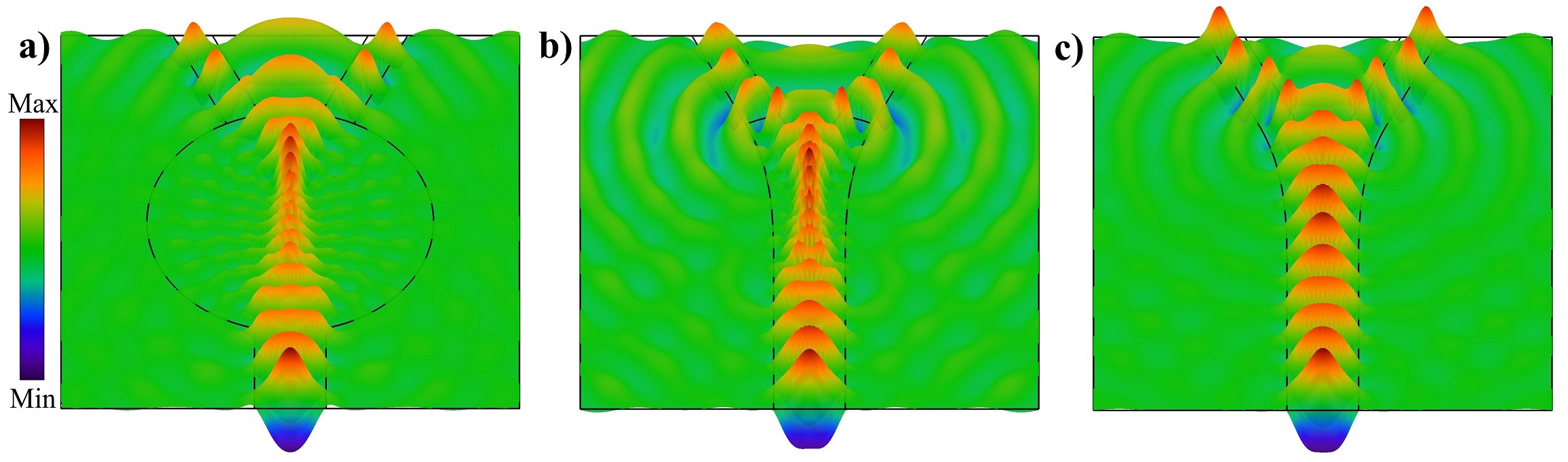}
\caption{Full-wave simulation results at the wavelength of $1550 nm$ for the splitting angle of $25^{\circ}$ with the a) complete GMFE lens, b) truncated GMFE lens, and c) without lens.}
\label{fig:Full}
\end{figure}

In this section, the 2D finite element method (FEM) simulations of the designed structures for the TE mode are presented at the wavelength of $1550 nm$. We evaluate the performance of the proposed splitters based on their splitting efficiency $(P_{split} / P_{in})$. $P_{split}$ is the power that transmits through each branching waveguide while $P_{in}$ is the power entering the splitting structure. For the splitter with $m=0.95$, where the branching angle is $25^{\circ}$, the full-wave simulation results of the complete and truncated lenses are displayed in Fig~\ref{fig:Full}. The splitter based on the complete lens with splitting efficiency of $32\%$ is shown in Fig~\ref{fig:Full}(a). The splitting efficiency of the truncated lens displayed in Fig~\ref{fig:Full}(b) is $33\%$. We also investigate the performance of the splitting structure without the lens in Fig~\ref{fig:Full}(c). In this case, the splitting efficiency is $35\%$ which is slightly better compared to the splitting structures of complete and truncated lenses. In low branching angles, the splitting structures based on lenses are not effective compared to the simple Y-junctions.

\begin{figure}[!t]
\centering
\includegraphics[width=1.0\linewidth]{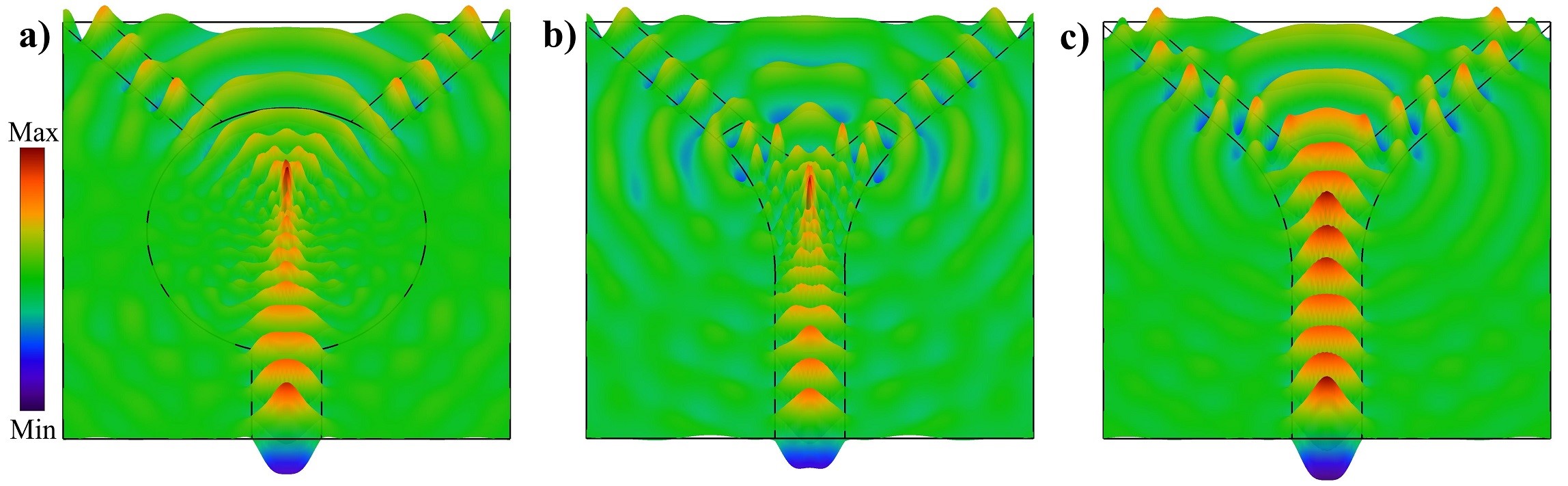}
\caption{Full-wave simulation results at the wavelength of $1550 nm$ for the splitting angle of $45^{\circ}$ with the a) complete GMFE lens, b) truncated GMFE lens, and c) without lens}
\label{fig:Full45}
\end{figure}

For the splitting structure with $m=0.87$, the results of the full-wave simulations are shown in Fig~\ref{fig:Full45}. In this design, the simulation results of complete and truncated lenses with the branching angle of $45^{\circ}$ are displayed in Fig~\ref{fig:Full45}(a) and Fig~\ref{fig:Full45}(b), respectively. The splitting efficiency of $29\%$ is achieved for the complete lens of Fig~\ref{fig:Full45}(a). The truncated lens of Fig~\ref{fig:Full45}(b) has the splitting efficiency of $22\%$. Wave propagation and the splitting performance of the structure without lens is shown in Fig~\ref{fig:Full45}(c). In this case, due to the fact that the optical signal is converted to higher mode, the splitting efficiency is $8\%$. As expected, the performance of the Y-junction without lens degrades considerably as the branching angle increases.

\begin{figure}[!t]
\centering
\includegraphics[width=1.0\linewidth]{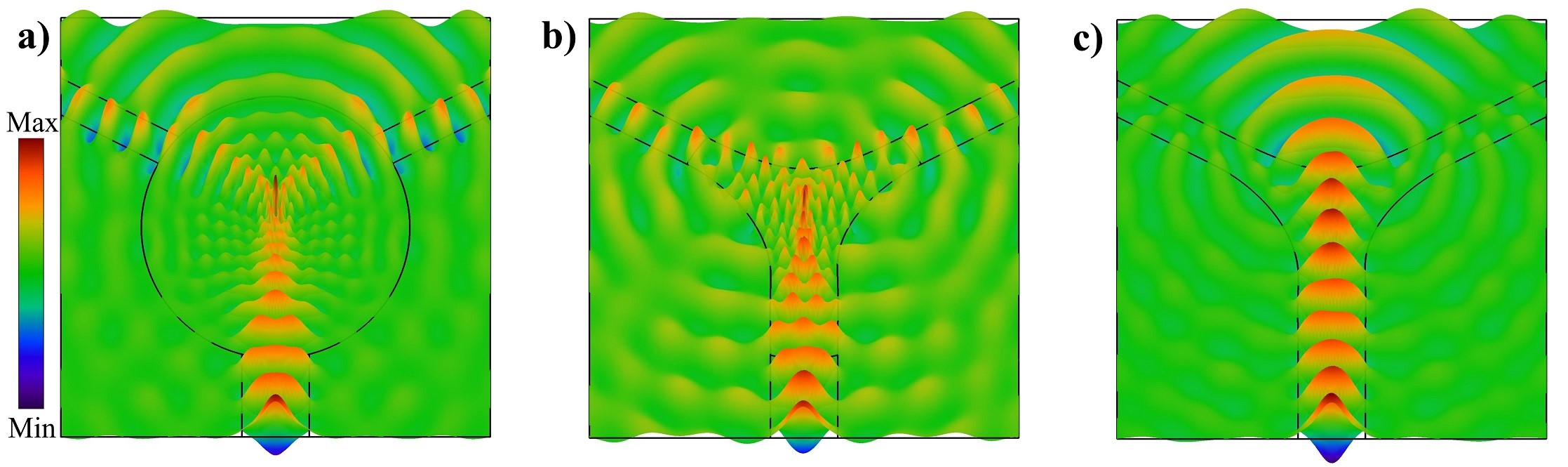}
\caption{Full-wave simulation results at the wavelength of $1550 nm$ for the splitting angle of $65^{\circ}$ with the a) complete GMFE lens, b) truncated GMFE lens, and c) without lens}
\label{fig:Fullsixty}
\end{figure}

For the splitter of Fig~\ref{fig:Rays_truncated}(c), the results of simulations are shown in Fig~\ref{fig:Fullsixty}. This structure comes with $m=0.80$ and a branching angle of $65^{\circ}$. Simulations for both the complete and truncated lenses are carried out. The splitter based on the complete lens with splitting efficiency of $35\%$ is shown in Fig~\ref{fig:Fullsixty}(a). For the truncated lens, shown in Fig~\ref{fig:Fullsixty}(b), the efficiency is $29\%$. We also investigate the performance of the splitting structure without the lens in Fig~\ref{fig:Fullsixty}(c) where the splitting efficiency is $0.2\%$. The performance of splitting structure without lens decreases meaningfully because the light wave propagates in a straight trajectory due to the lack of a suitable splitting structure.

\section{\label{Eaton}Eaton lens as splitter }
The Eaton lens can bend the light wave’s trajectory by $90^{\circ}$, $180^{\circ}$, or $360^{\circ}$. For a $90^{\circ}$ bend, the refractive index of the Eaton lens is given by \cite{badri2019low,badri2019polymer,lei2017generalized,tyc2011absolute,eskandari2019elliptical,du20163}

\begin{equation}
\label{eq:Eaton_n}
\begin{split}
 & {n_{lens}^2 = \frac{R_{lens}}{n_{lens}r} + \sqrt {{\left( {\frac{R_{lens}}{n_{lens}r}} \right)}^2 - 1} }
\end{split}
%
%
%
\end{equation}

\begin{figure}[!t]
\centering
\includegraphics[width=1.0\linewidth]{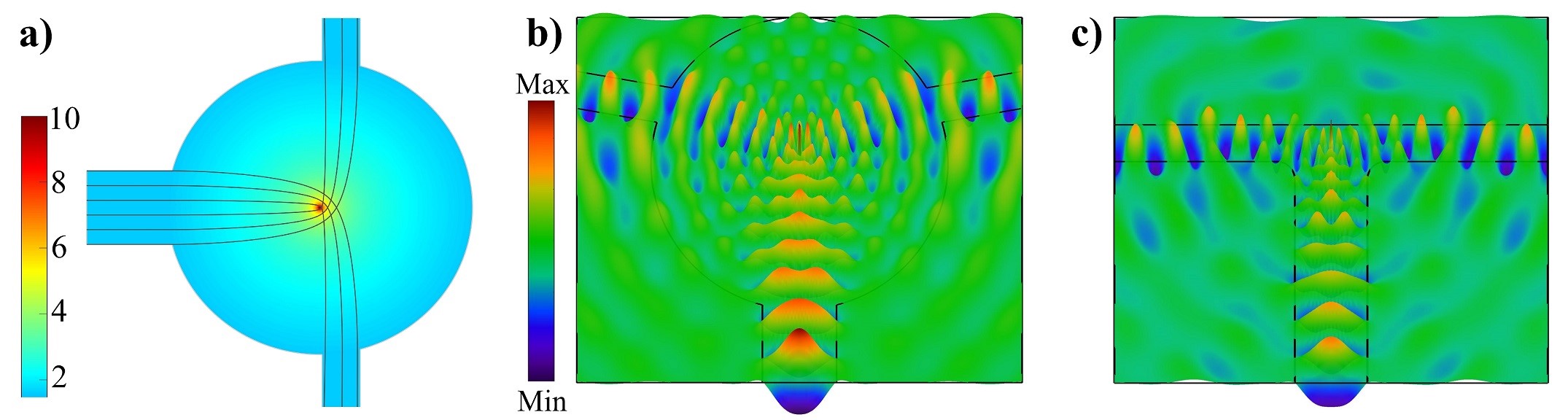}
\caption{Eaton lens as T-junction. a) ray trajectories for the Eaton lens. Full-wave simulation result at the wavelength of $1550$ nm for the b) complete lens and c) truncated lens.}
\label{fig:Fig7}
\end{figure}

The refractive index of the lens ranges from unity at its edge to infinity at the center of the lens. Minimizing the insertion loss is achieved by matching the refractive indices at the interface of the waveguides and the lens, thus the calculated   is multiplied by $n_{core}$. When an off-center beam is incident on the Eaton lens, power beam propagates inside the lens along a $90^{\circ}$ bending path which can be used to design waveguide bends \cite{badri2019low,badri2019polymer}. On the other hand, when an on-center beam is incident on the lens, it behaves like a T-junction \cite{yin2014all}. As shown in Fig~\ref{fig:Fig7}(a), an array of point sources covers a width of the waveguide core in ray-tracing calculations. Full-wave simulation reveals that for the splitting structure with the complete lens, the maximum splitting efficiency of $30\%$ is achieved when the branching angle is $80^{\circ}$. However, for a branching angle of $90^{\circ}$ with the truncated lens, the splitting efficiency of $41\%$ is achieved.

GRIN lenses have been implemented by graded photonic crystals \cite{gilarlue2018photonic,badri2019maxwell}, multilayer structures \cite{gilarlue2019multilayered}, varying the guiding layer thickness \cite{badri2019polygonal}. However, the proposed splitters cannot be implemented by these methods due to the extreme values at the center of the lenses and isotropic metamaterials should be used to implement these splitters.

\section{\label{Conc}Conclusion }
In summary, we investigated the possibility of designing power splitters for moderate index-contrast waveguides based on GRIN lenses. We designed splitters with branching angles of 25$^{\circ}$, 45$^{\circ}$, and 65$^{\circ}$ based on complete and truncated GMFE lenses. We also truncated the Eaton lens to design a 90$^{\circ}$ splitter. The GRIN lenses help to achieve a wide range of branching angles, however, this flexibility comes at the price of large values of refractive indices at the center of the lenses. The designed splitting structures cannot be implemented by conventional methods and they should be implemented by isotropic metamaterials. The advantage of the proposed designs compared to the TO-based splitters is the use of isotropic metamaterials instead of anisotropic metamaterials.

\bigskip
\bibliography{our}

\end{document}